\documentstyle[aps,prd,amstex,psfig]{revtex}

\newcommand\D{\widehat{D}}
\newcommand\order[1] { ${{\cal O}\! \left( #1 \right)}$ }
\newcommand{\eq}[1]{Eq.\ (\ref{#1})}

\newcommand{\lmax}{{\sf {L}}}

\newcommand{\etal}{{\em et al.\ }}

\begin{document}

\title{Fast, Exact CMB Power Spectrum Estimation
for a Certain Class of Observational Strategies}
\author{Benjamin D.~Wandelt$^1$ and Frode K.~Hansen$^2$}
\address{${}^{1}$Department of Physics, Princeton University,
Princeton, NJ 08544, USA\\
${}^{2}$Max Planck Institut f\"ur Astrophysik, Garching bei M\"unchen, Germany}

\maketitle              
\begin{abstract}    
We describe a  class of observational strategies for probing the
anisotropies in the cosmic microwave background (CMB) where the instrument
scans on rings which can be combined into an n-torus, the {\em ring torus}.
This class has the remarkable property that it allows exact maximum likelihood
power spectrum estimation in of order $N^2$ operations (if the size of the data
set is $N$) under circumstances which would previously have made this analysis
intractable: correlated receiver noise, arbitrary asymmetric beam shapes and
far side lobes, non-uniform distribution of integration time on the sky and
partial sky coverage. This ease of computation gives us an important
theoretical tool for understanding the impact of instrumental effects on CMB
observables and hence for the design and analysis of the CMB observations of
the future. There are members of this class which closely approximate the MAP 
and Planck satellite missions. We present a numerical example 
where we apply our ring torus methods 
to a simulated data set from a CMB mission covering a 20 degree patch on the
sky to compute the maximum likelihood  estimate of the power spectrum $C_\ell$ 
with unprecedented efficiency.

\end{abstract}

\section{Introduction}
A major near-term objective in the field of Cosmology today is to
gain a detailed measurement and statistical understanding of the
anisotropies of the cosmic microwave background (CMB). While the
theory of primary CMB anisotropy is well-developed (see
\cite{review} for a review) and we are  facing a veritable flood
of data from a new generation of instruments and missions,
the single most limiting hurdle are the immense computational
challenges one has to overcome to analyze these
data\cite{borrill,Gorski,BCJK}. 

We will now outline the key problems involved. The goal is to derive two
things: an optimal sky map estimate and a set of power spectrum estimates
$\widehat{C}_\ell$ which, for a 
Gaussian CMB sky are a
sufficient statistic, highly informative about cosmological
parameters. In this paper we will focus on the $\widehat{C}_\ell$ estimation
problem, but the ideas we develop can be usefully employed for
map-making as well \cite{challinor}.

We assume we start from a best estimate of the sky map of the cosmic microwave
background fluctuations,  the  monopole and
dipole components have been 
removed, and that residual foregrounds  are negligible after excision
of foreground dominated regions.  The analysis problem is then one of
spatial covariance estimation and involves 
maximizing the likelihood given the data as a function of the
$\widehat{C}_\ell$. 
If the observed CMB anisotropy data is written as a vector ${\bf d}$ the
likelihood takes the form
\begin{equation}
{\cal L}(C_\ell\vert d)=\frac{\exp\left[{-\frac12 {\bf d}^T ({\bf S}(C_\ell)+{\bf N})^{-1}{\bf d}}\right]}{\sqrt{(2\pi)^{N_{d}} \vert ({\bf
S}(C_\ell)+{\bf N}) \vert}},
\label{likelihood}
\end{equation}
where the signal covariance matrix $\bf{S}$ is a function of the
$C_\ell$, ${\bf N}$ is the noise covariance matrix and $N_d$ is the
number of entries in 
the data vector ${\bf d}$. If ${\bf S+N}$ is a general matrix, evaluating
the inverse and determinant in 
this quantity  takes \order{N^3_d} operations. For currently available
data sets $N_d\sim 10^5$, which is expected to grow to $N_d\sim 10^7$
by the end of this decade.


Borrill \cite{borrill} presented a careful review of the 
computational tasks involved in solving this 
maximization problem in the general case, using state-of-the-art
numerical methods. The conclusion is that
in their present form these methods fail to be practical for forthcoming CMB
data sets. Given the major international effort currently underway to
observe the CMB (see, e.g.~\cite{tophat,map,planck}), finding a
solution to this problem is of paramount importance. 

The underlying reason for this failure is that the signal covariance
matrix ${\bf S}$ and noise covariance matrix 
${\bf N}$ are usually very differently structured. The data ${\bf d}$
are usually represented by a vector  containing 
the temperatures of pixels in a sky map and $N_d=N_{pix}$, the number
of observed pixels. Hence entries in $\bf{S}$ and
${\bf N}$ are the covariances between  pairs of pixels  on the
celestial sphere.  In  pixel space,
${\bf S}$
is  full, while ${\bf N}$  is ideally  sparse (for a well-controlled
experiment). In spherical harmonic space we are in the opposite
regime. If the CMB sky is isotropic  ${\bf S}$ would be diagonal for all-sky observations, while the form
of ${\bf N}$ depends on experimental details and observational
strategy of the mission and in 
general could be any covariance matrix at all. Therefore the quantity
which enters in the likelihood 
${\bf C\equiv S+N}$ is not sparse in any easily accessible
basis. Other basis sets have been suggested, such as cut sky harmonics \cite{cutsky} and signal to
noise eigenmodes \cite{bond,tegfisher}, but changing into these more general  bases
takes \order{N_{pix}^3} operation.

There have been several attempts to find approximate formulae for the
likelihood. Several methods are based on replacing the likelihood
conditioned on the data vector  with an approximate likelihood based
on the pseudo-$C_\ell$ (the power spectrum computed on the incomplete sky, possibly weighted by an
apodizing window)\cite{WandeltHivonGorski,MASTER}. These methods
are very fast, scaling as \order{N_{pix}^{3/2}}, unbiased, but
sub-optimal to a degree which depends on the size of the survey
area. A similar approach is to first compute an unbiased estimate of the
correlation function in pixel space from the raw data and then 
transform to obtain the power spectrum in \order{N_{pix}^2} operations\cite{szapudi}.
A hierarchical approximation of the likelihood\cite{dore} reduces the
computational scaling to \order{N_{pix}^2}. 
Other authors have  found minimum variance quadratic
estimators\cite{tegmark,BJK} but the calculations still require 
\order{N_{pix}^3} operations. 

There is only one known class of observational
strategies in which the 
(numerically) exact
maximum likelihood solution can be computed faster, namely in \order{N_{pix}^2}
operations \cite{OhSpergelHinshaw}. This performance depends on a number of 
assumptions: azimuthally symmetric coverage of an azimuthally
symmetric portion of the sky with an 
circularly symmetric beam and uncorrelated noise\footnote{This class contains the ``ideal
CMB experiment'' with
uniform full sky coverage, uncorrelated noise,  and symmetric beam as
a special case. For pixelisation schemes admitting fast transforms,
such as HEALPix \cite{healpix} this ideal case can be solved in
\order{N_{pix}^{3/2}} operations.}. For ease of reference we will call this class {\em azimuthal
strategies}. 

In this paper we discuss a 
second class of observational 
strategies, introduced in \cite{wandelt}, for which the  maximum
likelihood problem can be solved exactly\footnote{The analysis of
CMB data on a one-dimensional data set consisting of a single ring
was first discussed in \protect\cite{DGH}. The generalisation of this
case to correlated received noise and arbitrary beam patterns is contained as a
special case in the class we discuss here.}.
This class is interesting in that it is
the the first to admit fast
maximum likelihood estimators which 
can analyze CMB data in the presence of correlated noise,
non-uniform
distribution of integration time, beam distortions,  far side lobes,
and particular types of partial sky coverage, all without needing to
invoke approximations. Every forthcoming CMB datasets will be affected
by some subset of these issues.

Further interest for observational strategies in  this class derives
from the fact that they are natural choices from a practical point of
view.  Evidence for this is that current and future balloon and
space missions have chosen strategies which are based on scanning on
interlocked rings: Archeops scanned on rings, MAP's precessing
scan generates a 3-torus,  and the simplest of the proposed
scanning strategies  for Planck generates a 2-torus.

The key point is that for this class of observational strategies modeling   
observations in the time ordered domain simplifies the
problem greatly. In section \ref{whyTOD} we argue that several instrumental
effects are modelled  more easily in the time ordered data (TOD). 
To be able to
formulate the likelihood problem 
in terms of the TOD one needs to be able to compute the expected signal
covariance between two samples in the TOD. We achieve this using the 
Wandelt-G\'orski method for fast convolution on the
sphere \cite{WandeltGorski} which we briefly review in section
\ref{convolve}. 

Using the results from these geometrical ideas we go on to formulate 
the likelihood problem for the $C_\ell$ on the ring torus in section
\ref{ringtorus}. We show that the correlation matrices have a special
structure for a class of scanning strategies 
namely those where the 
TOD can be thought of as being wrapped on an
n-torus, the {\em ring torus}. It turns out that both ${\bf S}$ and ${\bf N}$ are
sparse. In fact, they are both block-diagonal
with identical patterns of 
blocks; hence ${\bf C}$ is also block diagonal.
The exact computational scaling of the evaluation of  ${\cal L}(C_\ell)$
depends  on the scanning strategy used, but if the ring torus is an
2-torus it takes only \order{N_d^2} operations. 

We then
present a numerical example where er apply  this method to simulated data in
section \ref{example}.
Finally, we discuss the application of our ideas to real world
missions and conclude with suggestions for future directions in
section \ref{conclusions}.

\section{The Case for Analyzing CMB Data in the Time-Ordered Domain}
\label{whyTOD}
In this section we will discuss features of realistic instruments and
observational strategies and show that they are more easily modeled
in the time ordered domain. 

\subsection{Correlated noise}
Current detectors 
have the feature that they add correlated noise to the
signal. It appears to be a good 
approximation to assume that this  noise is  stationary and
circulant. If this is the case
then it has a very simple correlation structure in Fourier space. 
But estimating the sky from the TOD projects these simple correlations
into very complicated noise correlations between separated pixels,
which are visible as striping in the maps.

\subsection{Non-uniform distribution of integration time}
For most missions, practical constraints on the scanning
strategy force the available integration time to be allotted 
non-uniformly over the sky. This leads to coupling of the noise in
spherical harmonic space, in addition to coupling due to partial sky
coverage (see paragraph D below). However, in the TOD,
integration time per sample is constant, by definition. 

\subsection{Beam distortions and far side lobes}
Microwaves are macroscopic, with wavelengths of order
$10^{-3}$ --- $10^{-2}$cm. Hence they will diffract around macroscopic
objects. This 
leads to distorted 
point spread functions and far side 
lobes. Simulations have to convolve these asymmetric beams with an
input sky with foreground signals varying over
many orders of 
magnitude from place to place. Motions of the instrument while a
sample is taken leads to anisotropic smearing of the beam. Analyses have to deconvolve the
observations to make accurate inferences about the underlying sky.
Artifacts due to asymmetric beams and beam smearing affect the correlations of the
signal in a raw, coadded sky map in a complicated way which depends on the scanning
strategy of the instrument. If the beam is short in a particular
direction, the data contains information on this short
scale. Deconvolution can in principle reveal additional structure. 

\subsection{Partial sky coverage}
The cosmological information is most visible in spherical harmonic
space. In a homogeneous and isotropic Universe, where the
perturbations are Gaussian, the power spectrum $C_\ell$ completely encodes
the statistical
information which is present in a map of fluctuations. Statistical
isotropy of the fluctuations around us also implies that the spherical
harmonics $Y_{\ell m}$ are the natural basis for representing the CMB
sky. But due to astrophysical foregrounds that need to be excised from the map or
partial sky coverage by the instrument this isotropy is 
broken in actual data sets. This introduces additional signal and
noise correlations in $Y_{\ell m}$ space, because there is a geometrical
coupling between   $Y_{\ell m}$ of different $\ell$ and $m$.

\subsection{Pixelization effects}
Representing the data as a pixelised map on the sky is an
invaluable and time-proven tool for visual analysis. However, it is not without
possible disadvantages either. Binning samples into discrete sky pixels
introduces an extra level of smoothing on the pixel scale and erases
information, for example the exact distribution of sampling locations.
Too large pixels will degrade the signal and erase 
information about the  beam shape ({\em cf} issue C). If
the distribution of samples on the sky is non-uniform choosing pixels too 
small will result in many missing pixels and hence irregularly shaped observation
regions.

\vspace{.8cm}
There are currently no  exact and fast methods that can deal with
all of these issues.  Yet, every forthcoming CMB dataset will be affected
by some subset of these issues. 

It is clear from A and B that the noise properties are much simpler
in the TOD than in a sky map. But what about the signal correlations?
What is the price of forfeiting the advantages of the sphere for
representing statistically isotropic signals? 
In the next section we will
demonstrate that one can devise scanning strategies which break only
part of the symmetries of the sphere and make the signal correlations
in the TOD as simple as the noise properties.  

\section{Modeling the Signal in the Time-Ordered Domain}
\label{convolve}
Recently, Wandelt and G\'orski \cite{WandeltGorski} have
introduced new methods for greatly speeding up convolutions of
arbitrary functions on the 
sphere. This reference contains a detailed description of methods which will 
enable future missions such as MAP and Planck to 
take beam imperfections into account without resorting to
approximations.  The algorithm is completely general and
can be applied to any kind of directional data, even tensor fields
\cite{CMFAWG}. We will now briefly review the ideas and quote the results.

Let us define more precisely what is meant by ``convolution on the
sphere''. Mathematically, the 
convolution of two functions on the sphere is a function over the
group of rotations SO(3), because one has to 
keep track of the full relative orientation of the beam to the sky,
not just the direction it points in. For each possible relative
orientation (element of SO(3)) the convolution returns one value. To
specify the relative orientation one can 
use the Euler 
angles  $\Phi_1,\Theta$ and
$\Phi_2$\footnote{Our Euler angle convention refers to active right
handed rotations of a physical body in a fixed coordinate system. The
coordinate axes stay in place under all rotations and the object
rotates around the $z$, $y$ and $z$ axes by $\Phi_1$, $\Theta$ and
$\Phi_2$, respectively, according to the right handed screw rule.}.
The convolved signal for each beam orientation $(\Phi_1,\Theta,\Phi_2)$
can then be written as
\begin{equation}
T^S(\Phi_2,\Theta,\Phi_1)= \int d\Omega_{\vec{\gamma}}\,
\left\lbrack\D(\Phi_2,\Theta,\Phi_1) b\right\rbrack\!\!(\vec{\gamma})^\ast s(\vec{\gamma}).
\label{eq:start}
\end{equation}
Here the integration is over all solid angles, $\D$
is the operator of finite rotations such that $\D  b$ is the
rotated beam, and the asterisk denotes complex conjugation.
At each point on the sphere, there is a ring of
different convolution results corresponding to all relative
orientations about this direction. This is a direct consequence of the
fact that the beam is not assumed to be azimuthally symmetric.

If $\lmax$ measures the larger of the inverse of the smallest length
scale of the sky or beam, the numerical evaluation of  the integral in 
\eq{eq:start} takes \order{\lmax^2} operations for each tuple
$(\Phi_1,\Theta,\Phi_2)$. To allow subsequent interpolation at
arbitrary locations it is sufficient to discretize each Euler 
angle into  \order{\lmax} points and thus we have \order{\lmax^3}
combinations of them.
As a result, the total computational cost for evaluating the convolution
using \eq{eq:start} scales as \order{\lmax^{{5}}}.

It turns out that by Fourier transforming the above equation on the
Euler angles and judiciously factorising the rotation operator
$\D(\Phi_2,\Theta,\Phi_1)$ as described in \cite{WandeltGorski} we can 
reduce this scaling to \order{\lmax^{{4}}} in general. One can further
reduce the scaling to \order{\lmax^3} if one is only interested in the
convolution over a path which consists of a ring of rings at
equal latitude around the sphere. We called  such
configurations {\em basic scan paths} in references
\cite{WandeltGorski,wandelt}.

We quote here the formula for the Fourier coefficients of the 
convolved signal on this ring torus using the notation of
\cite{WandeltGorski}. In order to do so we need to make use of the
angular momentum representation of the rotation operator $\D$. 
A simple explicit expression for the matrix elements
$D^l_{m m'}(\phi_2,\theta,\phi_1)$ can be
given. One can define a real function $d^l_{m m'}(\theta)$ such that
\begin{equation}
D^l_{m m'}(\phi_2,\theta,\phi_1)=  e^{-im\phi_2}d^l_{m m'}(\theta)e^{-im'\phi_1}
\label{eq:D}
\end{equation}
Thus the dependence of $D$ on the Euler angles $\phi_1$ and $\phi_2$ is only in
terms of complex exponentials. While explicit formulas for the
$d$-functions exist \cite{BrinkSatchler}, they are more conveniently  computed
numerically using their recursion properties \cite{Risbo}.

If the convolved signal on the ring torus is $T^S_{rp}$ with $r$
counting the ring number and $p$ the pixel number on the ring, then
its Fourier coefficients are
\begin{equation}
\widetilde{T}^S_{m\,m'}=\sum_{\ell} s_{\ell m} d^\ell_{mm'}(\theta_E)  X_{\ell m'}.
\label{eq:specialresult}
\end{equation}
Here the $s_{\ell m}$ are the spherical harmonic multipoles for the CMB
sky and $\theta_E$ fixes the latitude of the spin axis on the
sky. The quantity  
\begin{equation}
X_{\ell m}\equiv \sum_{M} d^\ell_{mM}(\theta)b_{\ell M}^\ast
\label{eq:precompute}
\end{equation}
is just the rotation  of the beam multipoles $b_{\ell m}$ for an
arbitrary beam by $\theta$, the opening angle of the scan
circle. $X_{\ell m}$ can be precomputed. Counting indices in 
\eq{eq:specialresult} confirms
that computing the convolution along a basic scan path takes only
\order{\lmax^3} operations.

\begin{figure}[t]
\begin{center}
\leavevmode
\psfig {file=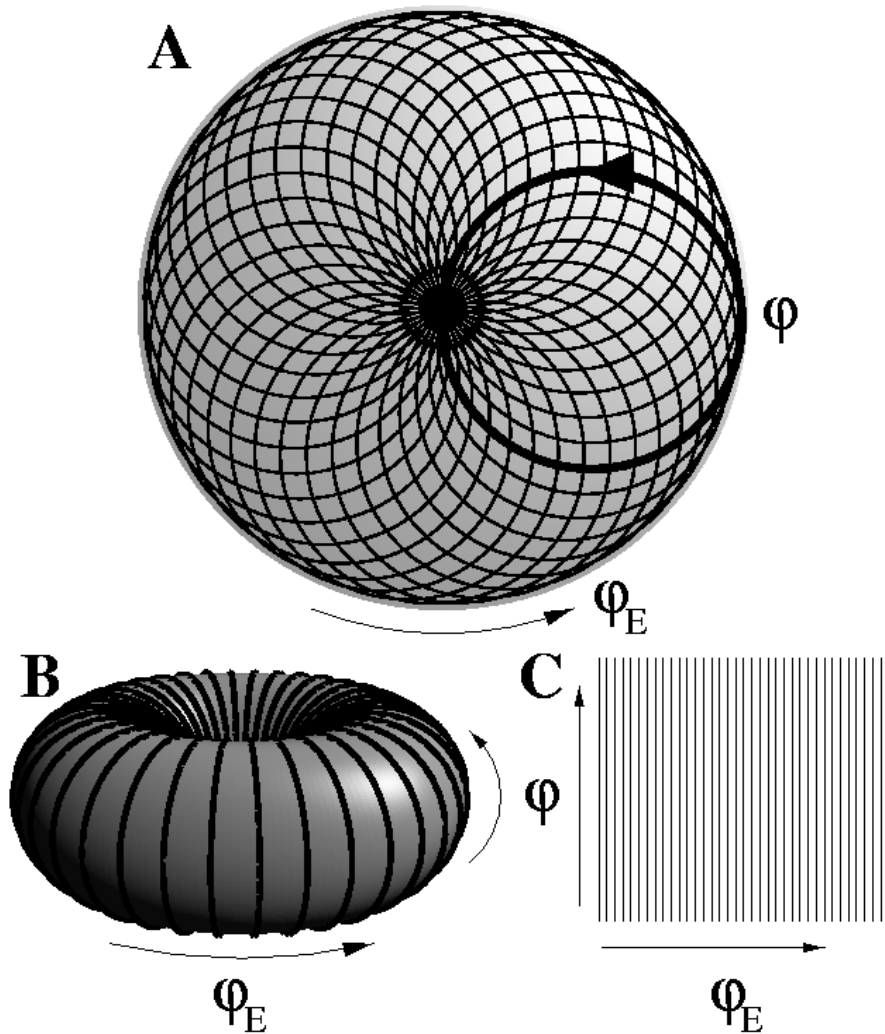,width=.7\textwidth}
\caption{The ring torus in three different representations. Figure A shows the
scan rings  on the sphere for one particular scanning strategy. We see
the projection of a disc $20^\circ$ across, covered with rings of
radius \protect{$\theta=5^\circ$} whose centers are
\protect{$\theta_E=5^\circ$} away from the north pole. The angles
\protect{$\phi$} and \protect{$\phi_E$} parametrize the position on
each ring and the ring location, respectively.  Figure B shows
the 2 dimensional ring torus generated by this scanning strategy and
in Figure C the rings are unfolded into a 2 dimensional plane like in
Figures (\ref{fig:ringsetnn}) and (\ref{fig:ringsetn}).}
\label{fig:toruscoords}
\end{center}
\end{figure}

\section{Likelihood Analysis on the Ring Torus}
\label{ringtorus}
Using the results from the previous section we can now write down a
statistical model in terms of the likelihood of the $C_\ell$ and experimental
parameters given the time-ordered data.

We will first derive the signal-correlation matrix  as a function of
the $C_\ell$ and then
go on to derive the noise correlation as a function of instrumental
parameters (like the shape of the noise power spectrum in the TOD) and
the scanning strategy (in this case the dimensions and position of the
basic scan path).

\subsection{Signal Correlations}

What we are interested in are the correlation properties of the signal
on the ring torus, \eq{eq:specialresult}. We can now easily compute
the correlation matrix  
\begin{eqnarray}
{\cal T}^S_{m\,m'\;M\,M'}&\equiv& \left\langle T^S_{m\,m'}
T^S_{M\,M'}\right\rangle\\
\label{eq:scor}
&=& \fbox{$\delta_{mM}$}\sum_{\ell} C_{\ell} d^\ell_{mm'}(\theta_E)  X_{\ell m'}d^\ell_{MM'}(\theta_E)
X_{\ell M'}.
\end{eqnarray}
where $\delta_{ij}$ denotes the Kronecker delta, defined as
\begin{equation}
\delta_{ij}=
\left\lbrace
\begin{array}{ll}
1 & {\rm if}~ i=j\\
0 & {\rm if}~ i\neq j\\
\end{array}\right..
\end{equation}
The boxed Kronecker delta shows that the  correlation matrix in
Fourier space on the ring torus is {\em block diagonal}. 

\subsection{Noise Correlations}

We start by assuming that we can model the noise in the time ordered
domain as a stationary
Gaussian process whose Fourier components $\widetilde{T}^N_k$ have the
property
\begin{equation}
\left\langle\widetilde{T}^N_k\widetilde{T}^{N\ast}_{k'}\right\rangle =
P(k)\delta_{kk'}.
\end{equation}

In the following we will allow for the complication that  the scanning
strategy involves integrating repeatedly on 
a fixed ring then the signal remains the same for each scan of
that ring. One can just co-add all scans over the
same ring without losing information. 

To be specific, if we define $N_R$ to be the number of pixels
per ring,  $N_c$  the
number of times the satellite spins while observing a fixed ring, and $N_r$
to be the number of rings for a one-year mission, we
can write $N_{TOD}= N_R N_c N_r$ for the number of samples in the
TOD.   In the following we will choose $N_r=N_R$ 
for simplicity, though this is not required by the formalism and can
be generalised trivially. 


To model the noise on the coadded rings, we define $T_{pr}$ to be the
noise temperature in the $p$th pixel in the $r$th ring, such that
\begin{equation}
T_{pr}\equiv \sum_{n=0}^{N-1} A_{pr \,n} T_n.
\end{equation} 
Here $A_{pr \,n}$ is  the {\em co-adding
matrix}, defined in \eq{coadd}. It encodes the 
scanning strategy by associating a given pixel in the ring set with
indices $r$ and $p$ with a set of samples in the TOD.

Then we can show  (see Appendix) that the covariance matrix of the 
Fourier components of the co-added noise on the rings is
\begin{eqnarray}
{\cal T}^N_{m'mM'M}&\equiv&
\left\langle\widetilde{T}^N_{m'm}\widetilde{T}^{N\ast}_{M'M} 
\right\rangle\label{noisecovariance}\\
&=&\frac{1}{(N_r)^3}\fbox{$\delta_{mM}$}\sum_{p,p'=0}^{N_r-1}e^{-\frac{2\pi
i}{N_r}  (m'p-M'p')} \sum_{\Delta =0}^{N_r-1}e^{-\frac{2\pi i}{N_r}  m
\Delta }C(\Delta ,p-p')  \nonumber
\end{eqnarray}
where $C(\Delta,p-p')$ is defined in \eq{first}.
The boxed Kronecker-$\delta$ is telling us that the noise
covariance is block diagonal in the ring Fourier basis.

\subsection{Generalization to more complicated scanning strategies}
\label{gen}
We just showed that by choosing the scanning strategy such
that the TOD can be wrapped (co-added) onto a  2-torus one can find a
basis  in which both noise and signal are block diagonal and hence
sparse. What is more, the data can be easily transformed into this basis
just by computing the Fast Fourier Transform (FFT) of the co-added TOD.

The results \eq{eq:scor} and \eq{noisecovariance} warrant some comments. We can
understand them in terms of symmetries. The ring
torus exhibits a spatio-temporal symmetry: time proceeds linearly
around the torus, 
in the same direction as the
azimuthal angle $\phi_E$ ({\em cf.}~Fig.~(\ref{fig:toruscoords})).
Wrapping the TOD on the torus breaks the translation invariance of the
noise in the TOD. Similarly, projecting the signal from the sphere onto
the torus breaks the isotropy of the signal on the sphere.
But these original symmetries manifest themselves as
ring--stationarity of the noise (correlations between rings only depend on the
distance between them) and ring--stationarity/``partial isotropy'' of the signal (only in
the azimuthal $\phi_E$ direction on the torus). These in turn express
themselves as a simplified correlation structure in Fourier space.

We can therefore see  that the key feature which leads to the
simplicity of the correlation matrices is the repetitive structure of
the scanning in the $\phi_E$ direction. Therefore it is plain
that these methods can be generalised to  ring tori of arbitrary
cross section, for example where the scanning rings are
ellipses, without losing the  computational scaling performance, albeit with
more cumbersome formulae which we will not produce here in detail.
This would also enable us to remove regions where foregrounds dominate
if that involved removing the same part of every ring in the
torus, because this does not spoil the ring--stationarity. 

From following the discussion in \cite{WandeltGorski} we can easily
see how the results  \eq{eq:scor} and \eq{noisecovariance} generalize
to precessing scans. For example rewriting Eq.~(8) in
\cite{WandeltGorski} as
\begin{equation}
 T_{m\,m'\,m''}=\sum_{\ell}
s_{\ell m} d^{\ell}_{m\,m'}(\theta_E) d^{\ell}_{m'm''}(\theta_P) X^{\;\ast}_{\ell m''}.
\label{eq:result}
\end{equation}
and substituting for $X_{\ell m''}$ from \eq{eq:precompute} we obtain the
signal part of the TOD for a precessing experiment where the spin axis
moves around the sphere at co-latitude $\theta_E$ and precesses about
it with amplitude $\theta_P$. We see that this TOD can be wrapped on a
3-torus. Computing the signal correlation matrix in Fourier space on this 3-torus
${\cal T}^S_{m\,m'\,m''\;M\,M',M''}$ we find that it is again block
diagonal. However, the size of blocks has now increased. The
computational performance in this case depends on the details of the
scanning strategy, but in general will lead to a  prefactor which
increases with the precession angle $\theta_P$.

For completeness we mention that by introducing further factors of  the rotation
operator this formalism can be generalised to n-tori with $n>3$, but
this would appear to give rise to impractical scanning strategies.

\subsection{Constructing the likelihood for ring torus power spectrum estimation}
One can write down the likelihood \eq{likelihood} on the ring torus
by simply substituting
\begin{equation}
d_{\rm map}\rightarrow d_{\rm ring~torus},\quad {\bf
S}\rightarrow{\cal T}^S,\quad
{\bf N}\rightarrow{\cal T}^N.
\end{equation}

Due to the block diagonality, the inverse and determinant in the
\eq{likelihood} 
only take \order{N^2} operations to 
evaluate. The gradient is similarly easy to evaluate
\begin{equation}
-2\frac{\partial\ln {\cal L}}{ \partial C_\ell}=
Tr({\bf C}^{-1} {\bf C,}_{C_\ell})- d^T{\bf C}^{-1} {\bf C,}_{C_\ell}{\bf C}^{-1}d
\end{equation}
which only takes \order{\lmax^3\sim N^{3/2}} operations to compute for
each component.

\subsection{Confidence Regions}
To determine confidence regions for the estimates one could explore
the likelihood surface around the 
maximum. If the number of estimated
band powers is $N_b$, using this recipe to evaluate the diagonal
elements of the covariance matrix of the estimates would scale as
\order{N_b N^2}, and the full covariance \order{N_b^2 N^2}. This is
more like \order{N^{5/2}} or \order{N^3} depending on the coarseness
of the binning.

But one can improve on this, because evaluating the gradient of the
likelihood also takes just \order{N^2} operations on the ring torus. So one can
numerically differentiate the gradient with respect to the $i$-th band power
estimate, to obtain the i-th row of the covariance matrix. The full
covariance matrix can then be computed in \order{N_b N^2}
operations. That's at most \order{N^{5/2}}  but given 
that the $C_\ell$ are smooth in most models $N_b\lesssim100$ may suffice,
especially for less than full sky coverage, where adjacent $C_\ell$
estimates become
strongly coupled.

\section{Numerical Example: Power Spectrum Estimation}
\label{example}
To test the method, we used a standard CDM power spectrum
and simulated a realization of spherical harmonic
multipoles $a_{\ell m}$ up to $\lmax=1024$. We used the Wandelt-G\'orski
method for fast convolution to generate $N_r=243$ scanning  rings with
  $5^\circ$ radius $N_R=243$ samples each. The spin axis was offset
from the north pole by $5^\circ$. This scanning pattern is shown in 
Fig.~(\ref{fig:toruscoords}A). The data covers a polar cap on the
sphere of diameter 20 degrees. For simplicity we used a symmetric
Gaussian beam with full width 
at half maximum of $18'$ -- however this simplification is not forced
by the method which allows using 
any beam pattern.

\begin{figure}[t]
\begin{center}
\leavevmode
\psfig {file=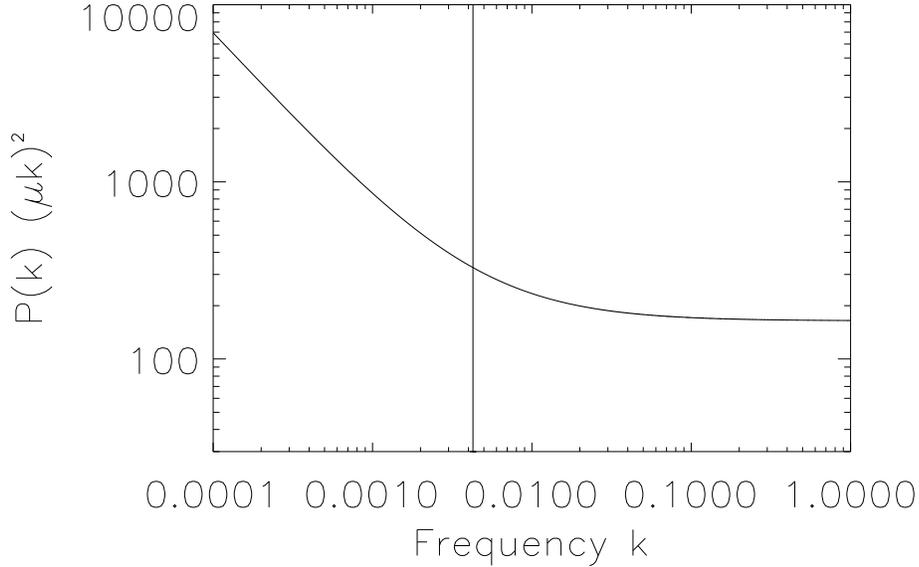,width=.7\textwidth}
\caption{The noise model in the time ordered data. The vertical line
shows the knee frequency. The frequency $k$ is shown in units of the
Nyquist frequency of the time stream $N_{TOD}/2$.}
\label{fig:noisepower}
\end{center}
\end{figure}

For the noise power spectrum we use a simple model motivated by
measurements from realistic receivers -- but again any power spectrum
could be used here. We adopot the functional form used by \cite{maino,delabrouille}
\begin{equation}
P(k)=2 \sigma^2[1+(k_{knee}/k)^\alpha],\qquad0<\alpha\leq2,
\label{pofk}
\end{equation}
where the first term corresponds to white noise with variance per
pixel of $\sigma^2$ and the second to $1/f$--type noise. 
We show this power spectrum in
Fig.~(\ref{fig:noisepower}). For our simulations we regularize this
form at small frequency by introducing a small additive constant
corresponding to a tenth of the smallest representable frequency  in the
denominator of the $1/f$ term. Hence our noise simulation is not
constrained to have zero mean but contains a random offset.

We generated a realisation of noise with $\alpha=1$, $\sigma=9\mu K$
and $k_{knee}=1.\times 10^{-3}$ (expressed in terms of the Nyquist
frequency $N_{TOD}/2$) and
added it to the ring set. The value of $\sigma$ was chosen
to give $S/N\sim 1$ at
$\ell=600$ for our simulations. We have $N_r=243$ rings with $N_R=243$
pixels in each ring. For simplicity of implementation each ring was
scanned a single time, $N_c=1$,  giving a total number  
of $N_{TOD}=59049$ pixels, a similar number of pixels to the BOOMERAnG
experiment.

\begin{figure}[t]
\begin{center}
\leavevmode
\psfig {file=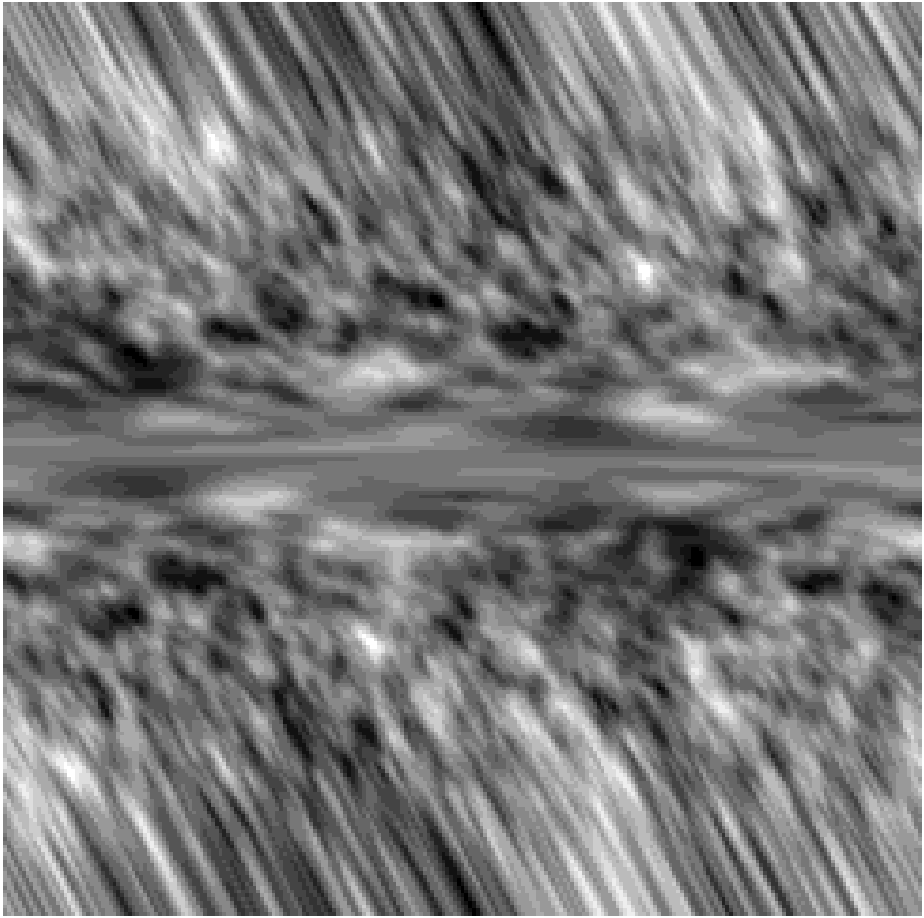,width=.7\textwidth}
\caption{The ring set from a scan on a simulated sky without 
noise. On the plot, a line from the top to the bottom
describes one ring. In this way the rings are put next to each other
from left ro right. The north pole gets mapped in to the horizontal
line of constant value in the middle of the plot. All  rings touch
there.}
\label{fig:ringsetnn}
\end{center}
\end{figure}

\begin{figure}[t]
\begin{center}
\leavevmode
\psfig {file=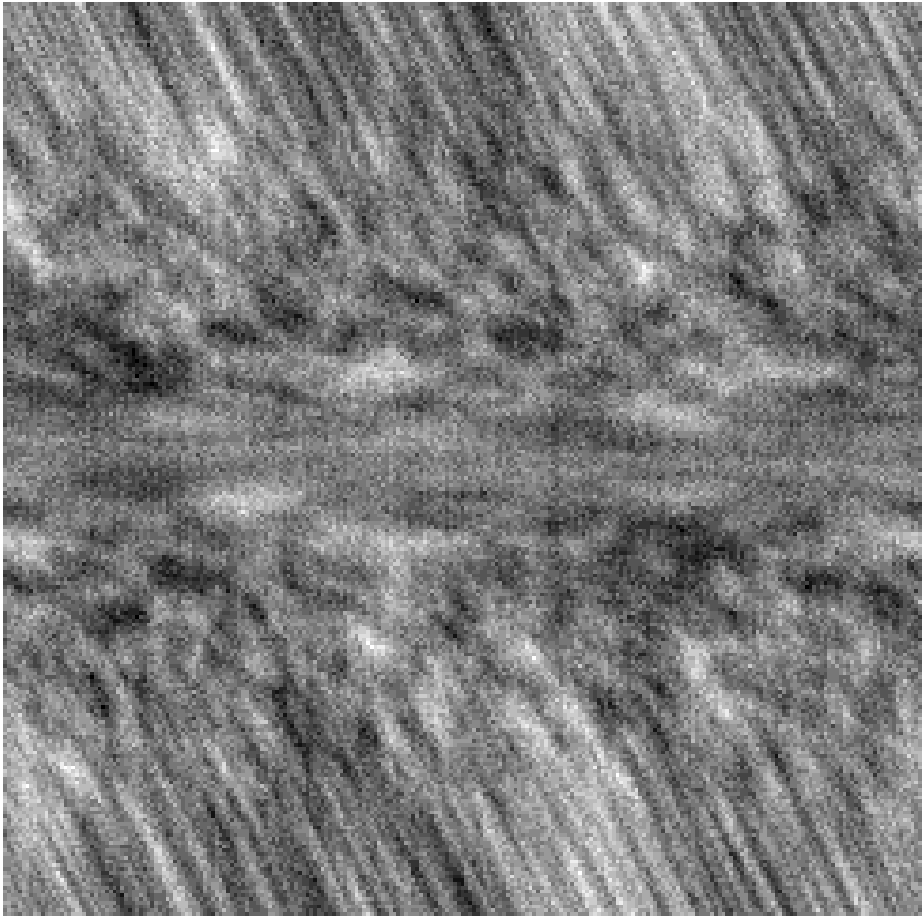 ,width=.7\textwidth}
\caption{As Fig.~(\ref{fig:ringsetnn}) but with correlated noise
added, causing the vertical striping.}
\label{fig:ringsetn}
\end{center}
\end{figure}

In Fig.~(\ref{fig:ringsetnn}) we show the ringset with and
in Fig.~(\ref{fig:ringsetn}) without noise using the representation
shown in Fig.~(\ref{fig:toruscoords}C). On the plots, a line from the top to the bottom
describes one ring, in this way the rings are put next to each other
from left ro right. The line from left to right in the middle, with
the same value, is the north pole where all rings meet. We see clearly
the stripes from the correlated noise.

Instead of estimating $C_\ell$, we estimate $D_\ell$ given by,
\begin{equation}
D_\ell=\frac{C_\ell\ell(\ell+1)}{e^{-\sigma_b^2\ell(\ell+1)}},
\end{equation}
where $\sigma_b$ is for a $18'$ FWHM beam. We divide $D_\ell$ in
$N_b=20$ flat bins $D^1, D^2, ..., D^{20}$ with $50$ multipoles in
each.


The Fourier transformed signal correlation matrix can be calculated in
two ways. One might evaluate it directly in Fourier space with equation
(\ref{eq:scor}), or one could first evaluate it in pixel space and then
Fourier transform it using FFT. In our test example we have chosen
the latter. The formula for the correlation matrix in pixel space is
in our case
\begin{equation}
<T_{pr}T_{p'r'}>=\sum_{b}D^b\sum_{\ell\epsilon
b}\frac{e^{-\sigma_b^2\ell(\ell+1)}}{\ell(\ell+1)}\frac{2\ell+1}{4\pi}P_\ell(\cos\delta\theta),
\end{equation}
where $\delta\theta$ is the angular distance between the two points
$pr$ and $p'r'$ on the sphere and the sum over $\ell$ is taken over
all $\ell$ values in the particular bin $b$. All $\delta\theta$ angles are precomputed. The
derivative with respect to a certain bin is then simply,
\begin{equation}
\frac{\partial <T_{pr}T_{p'r'}>}{\partial D^b}=\sum_{\ell\epsilon
b}\frac{e^{-\sigma_b^2\ell(\ell+1)}}{\ell(\ell+1)}\frac{2\ell+1}{4\pi}P_\ell(\cos\delta\theta),
\end{equation}

\begin{figure}[t]
\begin{center}
\leavevmode
\psfig {file=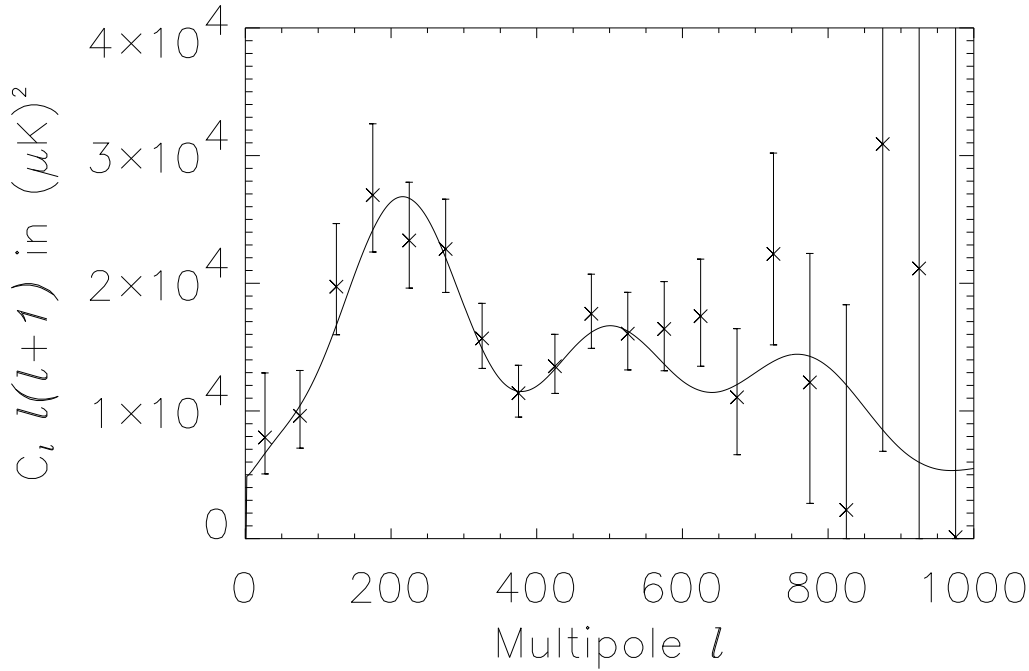,width=.8\textwidth}
\caption{The result from a likelihood estimation of the power spectrum
from a time stream with 243 rings with 243 pixel per ring. The
error bars approximate asymmetric $2\sigma$ confidence regions by showing where the
loglikelihood drops by less than $2$ compared to the maximum. Three
points at high $\ell$ fail to detect significant power at 95\% confidence
and return upper limits on $C_\ell$.} 
\label{fig:est}
\end{center}
\end{figure}

In Figure (\ref{fig:est}) we have plotted the result from
likelihood estimation. To find this estimate we initialised the
iterative conjugate gradient scheme \cite{NumRec} with a flat power
spectrum $C_\ell=$constant. The minimum of the likelihood was found
after about 30 function and derivative 
evaluation taking about 2 hours each on a single processor on a 500 MHz
DEC Alpha Work Station. If we had started the likelihood minimisation
with a better inital guess for the power spectrum (which could be
obtained using one of the $N^{3/2}$ methods \cite{WandeltHivonGorski,MASTER}) the
number of likelihood evaluations could be significantly reduced.

 The solid line shows the input average power spectrum used
and the crosses are the estimates with error bars. The
errorbars approximate a 95\% confidence region determined by 
finding the interval of the estimate where the
log-likelihood is reduced by less than 2 compared to its maximum value. 
We compared these confidence regions with the errors bars determined
from the observed Fisher matrix by finding the numerical second
derivative of the loglikelihood. 
The square root of the diagonal components of its inverse yields
errorbars which were consistent 
with the ones we obtained from the likelihood contours.

Note that we did not subtract lower order modes from the sky patch
that we simulated (though we assumed that the map was free of
non-cosmological monopole and dipole modes). As we mentioned, the noise 
was also not constrained to average to zero. The presence of lower
order modes was correctly modelled and handled by
the algorithm.  

\section{Critical Discussion and Conclusions}
\label{conclusions}
In this paper we described the properties of 
the class of observational stratgies where the instrument scans on
rings which can be arranged into an n-torus. Using these properties we 
constructed  for the first time an exact formulation of
the maximum likelihood problem of power spectrum analysis on the
sphere which can deal with correlated noise, realistic beams, partial
sky coverage, and non-uniform noise with a computational scaling of
\order{N^2}$\!\!$. This is a significant advance over other available exact
methods with the same scaling, all of which assume white noise and
azimuthally symmetric beams, amongst other assumptions.

How realistically can we expect these methods to be useful for real
CMB missions? Real observations obviously
never coincide exactly with any simple mathematical model. This was also true
for the previously known solvable class of  strategies of azimuthal sky
coverage, symmetric beams and uncorrelated noise. But the fact that
the likelihood simplifies in this idealised situation, defined one of
the design goals for the MAP satellite, namely to ensure that the
noise be uncorrelated between samples.
This allowed a fast maximum likelihood power spectrum estimation
method to be developed for MAP \cite{OhSpergelHinshaw} which achieves success
by employing iterative numerical schemes taking advantage of the fact that
the  data nearly satisfy the assumptions of the exactly solvable
case and thus improve convergence. Perturbation theory is powerful but
only if exact solutions are known that are close to the real thing.

In our case, even though the methods we describe can deal with
many features of actual CMB missions, there remain
discrepancies. We
have not shown (except for our discussion in section
\ref{gen}) how to deal with small areas where foregrounds
dominate and defy removal, such as bright dust emission from the galactic
plane,  or certain point sources.  
Also, irregularities
in the scanning strategy will somewhat spoil the symmetry of spatial
correlations in the signal and temporal correlations in the noise
which the ring torus exploits. 

We therefore advocate a similar  approach to the one
which  was used successfully in 
\cite{OhSpergelHinshaw}, namely to use exact inverses, computed using
the methods we presented in this  
paper as preconditioners to speed up conjugate
gradient methods for computing $C^{-1}d$ for more general cases, e.g.
galactic cuts, point sources, imperfect scanning and non-stationarity
of the noise. Other nuisance factors, such as discrete events
like cosmic ray hits and short telemetry losses can plausibly be dealt
with in established ways, namely by filling in the data stream with a
constrained realisation based on the properties of the observed data.

Even neglecting the computational advantages  of ring torus methods
there are other advantages to analysing time ordered data directly.
First, this approach does not rely on  map-making on the sphere for the purposes
of power spectrum estimation.  Not having to base $C_\ell$ estimation
on maps helps to  avoid systematic effects or signal degradation
introduced by previous analysis steps, for example a heuristic
de-striping tool.
On the ring torus, one fully models the striping due to 
correlated noise and hence does not rely on heuristic methods. 

While the sky map may not necessarily be the best
starting point for the purposes of $C_\ell$ estimation, one should
of course make the best map one possibly can. Maps are indispensable
for the visualization of the data and  identifying systematic
errors. Also, recently developed map based methods 
for estimating the noise properties 
of the observations from the data itself
\cite{FerreiraJaffe,Prunet} can be used to provide the necessary noise model
input to our methods (such as $P(k)$, \eq{pofk}).

In fact the geometrical ideas we discuss in this paper may well be
useful for map making itself. Reference \cite{WandeltGorski} shows 
how the ring torus speeds up iterative methods for deconvolving a map with
asymmetric beams. The block diagonal noise correlations on the torus
mean that one can use the same geometrical ideas for solving the map
making equations in the presence of noise.
In fact both signal and noise
covariances are sparse in this basis, hence
this method can be used to quickly Wiener filter the data on the ring torus
once the $C_\ell$ are known to optimally estimate the CMB
fluctuations on the torus. These can then be visualized by projecting
them onto a sky map.

Under the stated assumptions we have shown that the power spectrum
analysis problem simplifies 
greatly when it is considered in the domain of co-added time-ordered
data which has the geometry of a torus. We indicate how to derive
extensions of the formulae we presented here to more 
general scanning strategies (e.g. a precessing scan, such as MAP's or
another proposal for Planck). In this case the efficiency depends on the
size of the extra dimensions which the torus acquires. 

However, even as they stand, the methods and ideas we describe
represent a new theoretical tool which allows evaluating the impact of
CMB mission design choices in much 
more detail than previously feasible. In order to achieve the
best possible estimates of the CMB power 
spectrum, or certain cosmological parameters, one can ask whether it
is more important to reduce the amount of correlated 
noise from the receiver, to ensure accurate pointing, to increase sky
coverage, change the survey geometry, or to control the beam shape and
far side lobes?  We are investigating quantitative answers to these questions.

\acknowledgements
We would like to thank A.~J.~Banday, A.~Challinor, K.~M.~G\'orski,
E.~Hivon,  D.~Mortlock and
members of the Planck collaboration for discussions. We acknowledge
the use of the HEALPix \cite{healpix} and CMBFAST 
\cite{cmbfast} packages in preparing this 
publication. BDW is supported by the NASA MAP/MIDEX
program. FKH is supported by a grant from the Norwegian Research Council.

\appendix
\section{The Noise Covariance on the ring torus}

The co-adding matrix can be written  as
\begin{equation}
A_{pr\,n}\equiv\frac{1}{N_c}\delta_{p(n \mod
N_r)}\delta_{r\left(\left\lfloor\frac{n}{N_r
N_c}\right\rfloor\right)}=
\frac{1}{N_c}\delta_{rx}\delta_{pz}
\label{coadd}
\end{equation}
Here $\lfloor f \rfloor$ is defined to be the largest integer smaller
than $f$.

The previous equations simplify if we
write $n$ as $n=N_r N_c x+N_r y+z$ with 
$x\in[0,N_r-1]$, $y\in[0,N_c-1]$, $z\in[0,N_r-1]$.
Now using that
\[
<T^N_nT^N_{n'}>=\sum_ke^{\frac{-2\pi ik}{N}(n-n')}P(k),
\] 
we find that the noise covariance on the ring set is
\begin{equation}
C(r-r',p-p')\equiv \left\langle{T}^N_{rp}{T}^{N\ast}_{r'p'}\right\rangle=\frac{1}{N_c^2}\sum_{k=0}^{N-1}P(k)\sum_{y,y'=0}^{N_c-1}
e^{{-\frac{2\pi i k}{ N}(N_r N_c (r-r')+N_r(y-y') + (p-p'))}}.
\label{first}
\end{equation}
Note that $C$ is periodic with period $N_r$ in its first entry;
letting $\Delta\equiv r-r'$ we have
\begin{equation}
C(\Delta ,p-p')=C(\Delta +j N_r,p-p'), \quad j\in {\cal Z}.
\label{circulant}
\end{equation}

Finally we Fourier transform to arrive at the covariance of the
Fourier components of the 
the co-added noise on the rings
\begin{eqnarray}
{\cal T}^N_{m'mM'M}&\equiv&
\left\langle\widetilde{T}^N_{m'm}\widetilde{T}^{N\ast}_{M'M} 
\right\rangle\\
&=&\frac{1}{(N_r)^3}\fbox{$\delta_{mM}$}\sum_{p,p'=0}^{N_r-1}e^{-\frac{2\pi
i}{N_r}  (m'p-M'p')} \sum_{\Delta =0}^{N_r-1}e^{-\frac{2\pi i}{N_r}  m
\Delta }C(\Delta ,p-p')  \nonumber
\end{eqnarray}
where we used (\ref{circulant}) to simplify the sums over $r,r'$ and to
obtain the boxed Kronecker-$\delta$ telling us that the noise
covariance is block diagonal in the ring Fourier basis.

For practical purposes one would first compute the Fourier transfrom
in (\ref{first}) using an FFT taking $\sim N\log N$ operations. Then
the resulting array can be interpolated at $N_r(2N_r-1)$ positions to
obtain the independent components of $C$. Finally we have a
three-dimensional FFT left to evaluate (\ref{noisecovariance}), which
requires $N_r^3 \log N_r$ operations.

\end{document}